\documentclass[epj]{svjour}

\usepackage{graphicx}
\usepackage{dcolumn}
\usepackage{bm}
\usepackage{graphics}
\usepackage{epsfig,color}

\newcommand{\be}{\begin{equation}}
\newcommand{\ee}{\end{equation}}
\newcommand{\bea}{\begin{eqnarray}}
\newcommand{\eea}{\end{eqnarray}}
\newcommand{\la}{\langle}
\newcommand{\ra}{\rangle}

\newcommand{\bS}{\bf S}

\begin{document}
\title{Magnetic phase diagram of the dimerized spin $S=\frac{1}{2}$ ladder}
\titlerunning{Magnetic phase diagram of the dimerized spin $S=\frac{1}{2}$ ladder}
\author{G.I. Japaridze$^{1}$  and
S. Mahdavifar$^{2}$}

\authorrunning{G.I. Japaridze and S. Mahdavifar}
%
%
\institute{$^{1}$ Institut f\"ur Theoretische Physik, Universit\"at zu K\"oln, Z\"ulpicher Str. 77,
D-50937 K\"oln, Germany \\
\hspace*{2mm}  and Andronikashvili Institute of Physics, Tamarashvili 6, 0177, Tbilisi, Georgia\\
$^{2}$ Department of Physics, University of Guilan, 41335-1914, Rasht, Iran}

\date{Received: \today }

\abstract{The ground-state magnetic phase diagram of a spin $S=1/2$ two-leg
ladder with alternating rung exchange $J_{\perp}(n)=J_{\perp}[1 +
(-1)^{n} \delta]$  is studied using the analytical and numerical
approaches. In the limit where the rung exchange is dominant, we
have mapped the model onto the effective quantum sine-Gordon model
with topological term and identified two quantum phase transitions
at magnetization equal to the half of saturation value from a
gapped to the gapless regime. These quantum transitions belong to
the universality class of the commensurate-incommensurate phase
transition. We have also shown that the magnetization curve of
the system exhibits a plateau at magnetization equal to the half
of the saturation value. We also present a detailed numerical analysis of the low energy
excitation spectrum and the ground state magnetic phase diagram of
the ladder with rung-exchange alternation using Lanczos method of
numerical diagonalizations for ladders with number of sites up to $N=28$. We have
calculated numerically the magnetic field dependence of the
low-energy excitation spectrum, magnetization and the on-rung
spin-spin correlation function. We have also calculated the width
of the magnetization plateau and show that it scales as
$\delta^{\nu}$, where critical exponent varies from $\nu
=0.87\pm0.01$ in the case of a ladder with isotropic
antiferromagnetic legs to $\nu =1.82\pm0.01 $ in the case of
ladder with ferromagnetic legs. Obtained numerical results are in
an complete agreement with estimations made within the
continuum-limit approach.}


\maketitle

\section{Introduction}\label{sec1}

Low-dimensional quantum magnetism has been the subject of intense
research activity since the pioneering paper by
Bethe \cite{Bethe_31}. Perpetual during the decades interest in study
of these systems is determined by their remarkably rich and
unconventional low-energy properties (see for
review Ref. 2). An increased current activity in this field is
connected with the large number of qualitatively new and
dominated by the quantum effects phenomena discovered in these
systems \cite{Suchdev_NP_08,Giamarchi_NP_08} as well as with the
opened wide perspectives for use low-dimensional magnetic
materials in modern nanoscale technologies.

The spin $S=1/2$ two-leg ladders represent one, particular subclass
of low-dimensional quantum magnets which also has attracted a lot of
interest for a number of reasons. On the one hand, there was
remarkable progress in recent years in the fabrication of such
ladder compounds \cite{RiceDagotto}. On the other hand, spin-ladder
models pose interesting theoretical problems since antiferromagnetic
two-leg ladder systems with spin $S=1/2$ have a gap in the
excitation spectrum \cite{Schulz_86,Affleck_91,Nersesyan_96} they
reveal an extremely rich behavior, dominated by quantum effects in
the presence of a magnetic field. These quantum phase transitions
were intensively investigated both theoretically [9-24] and
experimentally [25-30].

Usually, these most exciting properties of low dimensional quantum
spin systems exhibit strongly correlated effects driving them
toward regimes with no classical analog. Properties of the
systems in these regimes or "quantum phases" depend in turn on
the properties of their ground state and low-lying energy
excitations. Therefore search for the gapped phases emerging from
different sources and study of ordered phases and quantum phase
transitions associated with the dynamical generation of new gaps
is an important direction in theoretical studies of quantum spin
systems.

A particular realization of such scenario appears in the case where
the spin-exchange coupling constants are spatially modulated. The
spin-Peierls effect in spin chains represent prototype example of
such behavior \cite{Gross_Fisher_79}. In the recent paper of one of
the authors the new type of effective spin-Peierls phenomenon in
ladder systems, connected with spontaneous dimerization of the
system during the magnetization process via alternation of rung
exchange has been discussed \cite{JP_06}. The Hamiltonian of the
model is
\begin{eqnarray}
{\cal H} &=& J_{\parallel} \sum_{n,\alpha} {\bS}_{n,\alpha} \cdot
{\bS}_{n+1,\alpha} - H  \sum_{n,\alpha}
S^{z}_{n,\alpha} \nonumber\\
&+& J_{\perp} \sum_{n} \left[1 + (-1)^{n} \delta
\right] {\bS}_{n,1} \cdot {\bS}_{n,2} \, , \label{L_w_ARE_Hamiltonian}
\end{eqnarray}
where $\bS_{n,\alpha}$ is the spin $S=1/2$ operator of rung n
(n=1,...,L) and leg $\alpha$ ($\alpha=1,2$). The interleg coupling
is antiferromagnetic, $J^{\pm}_{\perp} = J_{\perp}(1 \pm \delta)
> 0 $.

In  Ref. 32 the model was studied analytically in the limit of
strong rung exchange and magnetic field $J^{\pm}_{\perp} \simeq H
\gg |J_{\parallel}|,\delta J_{\perp}$. In this limit the ladder
Hamiltonian is mapped onto the spin-1/2 $XXZ$ Heisenberg chain in
the presence of both longitudinal uniform and staggered magnetic
fields, with the amplitude of the staggered component of the
magnetic field proportional to $\sim \delta J^{0}_{\perp}$. The
continuum-limit bosonization analysis of the effective spin-chain
Hamiltonian show, that the alternation of the rung-exchange leads to
the dynamical generation of a new energy scale in the system and to
the appearance of two additional quantum phase transitions in the
magnetic ground state phase diagram. These transitions manifest
themselves most clearly in the presence of a new magnetization
plateau at magnetization equal to one half of its saturation value.
It was shown that the new commensurability gap (correspondingly the
width of magnetization plateau) scales as $\delta^{\nu}$, where $\nu
=4/5$ in the case of a ladder with isotropic antiferromagnetic legs
and $\nu =2$ in the case of a ladder with isotropic ferromagnetic
legs. Although up to now no materials are available
which realize this models, theoretical studies of this type of
Peierls distortion is extremely interesting, since such an
instability could in principle develop during the magnetization
process of an antiferromagnetic ladder, in particular with high
probability in the case of applied uniaxial along legs pressure.

In this paper we continue our studies of an isotropic spin $S=1/2$
two-leg ladder with alternating rung exchange in a uniform magnetic
field using the numerical analysis. In particular we apply the
Lanczos method to diagonalize numerically finite ladder systems with
lengths $L=6, 8, 10, 12, 14$. Using the exact diagonalization
results, we calculate the spin gap, magnetization and the intra-rung
spin correlations as a function of applied magnetic field. We have
also calculated the spin-density distribution in the ground state at
magnetization plateau. Based on the exact diagonalization results we
obtain the ground-state magnetic phase diagram of the model showing
four quantum phase transitions, in agreement with the predictions
made in Ref. 32. We also numerically computed the width of the
plateau and show it scales as $\delta^{\nu}$, where
$\nu=0.87\pm0.01$ in the case of a ladder with isotropic
antiferromagnetic legs and $\nu=1.82\pm0.02$ in the case of a ladder
with isotropic ferromagnetic legs.

The paper is organized as follows. In the forthcoming section we
briefly summarize results of the analytical studies. In section
III, we present numerical results of our exact diagonalization
studies of the system. Finally, we conclude and summarize our
results in section IV.

\section{Effective Hamiltonian}\label{sec2}

In this section we briefly summarize the results obtained within
the analytical studies. To obtain the effective spin-chain we
follow the route already used to studies of the standard two-leg ladder
models in the same limit of strong rung
exchange \cite{Mila_98,Totsuka_98}. We start from the case
$J_{\parallel} = 0$, where the system decouples into a set of
noninteracting rungs with couplings $J_{\perp}(n)=J_{\perp}[1\pm
(-1)^{n}\delta]$ and an eigenstate of the Hamiltonian is written
as a product of rung states. At each rung, two spins form either a
singlet state $|s\ra$ with energy $E_{s}(n) = -0.75J_{\perp}(n)$
or in one of the triplet states $|t^{+}\ra$,$|t^{0}\ra$ and
$|t^{-}\ra$ with energies $E_{t^{+}}(n) = 0.25J_{\perp}(n) - H$,
$E_{t^{0}}(n) = 0.25J_{\perp}(n)$ and
$E_{t^{-}}(n)=0.25J_{\perp}(n) + H$, respectively. When $H$ is
small, the ground state consists of a product of rung singlets.
As the field $H$ increases, the energy of the triplet state
$|t^{+}\rangle$ decreases and at $H \simeq J_{\perp}(n)$ forms,
together with the singlet state, a doublet of almost degenerate
low energy state, split from the remaining high energy two
triplet states. This allows to introduce the effective spin
operator $\tau$ which act on these states as \cite{Mila_98}
\begin{eqnarray}
&&\tau_{n}^{z}|\,s_{0}>_{n}~ = -\frac{1}{2}|\,s_{0}>_{n}\, , ~~~~
\tau_{n}^{z}|\,t^{+}>_{n} ~ = \frac{1}{2}|t^{+}>_{n}\, ,\nonumber \\
&&\tau_{n}^{+}|\,s_{0}>_{n} ~ = ~~~|\,t^{+}>_{n}\, , ~~~~~
\tau_{n}^{+}|t^{+}>_{n}~ = ~~~0 \, , \\
&&\tau_{n}^{-}|\,s_{0}>_{n} ~ = ~~~~ 0 \, ,~~~~~~~~~~~~
\tau_{n}^{-}|\,t^{+}>_{n} = |\,s_{0}>_{n}\, . \nonumber
\end{eqnarray}
The relation between the real spin operator ${\bS}_{n}$ and the
pseudo-spin operator ${\mbox{\boldmath $\tau$}}_{n}$ in this
restricted subspace can be easily derived by inspection,
\be S^{\pm}_{n,\alpha} =
(-1)^{\alpha}\frac{1}{\sqrt{2}}\tau^{\pm}_{n}\, , \quad
S^{z}_{n,\alpha} = \frac{1}{2}\left(\frac{1}{2}+\tau^{z}_{n}\right)
\, . \label{S-Tau-relations} \ee
Using (\ref{S-Tau-relations}), to the first order and up to a
constant, we easily obtain the effective Hamiltonian
\begin{eqnarray}
H_{eff}& = & \sum_{n} \{\frac{1}{2}J_{xy}\left(\tau_{n}^{+}
\tau_{n+1}^{-} + h.c.\right) + J_{z} \tau_{n}^{z}
\tau_{n+1}^{z}\}\nonumber\\ & - &
h_{eff}^{0}\sum_{n}\tau_{n}^{z}  -
h_{eff}^{1}\sum_{n}(-1)^{n}\tau_{n}^{z} \, ,
\label{L_w_ARE_Effective_XXZ_Hamiltonian}
\end{eqnarray}
where $J_{xy}  =  2J_{z}=J_{\parallel}$ and
\begin{eqnarray}
h_{eff}^{0} & = & H - J_{\perp} - J_{\parallel}/2\, , \qquad
h_{eff}^{1}  =  \delta J_{\perp}\, . \label{eff}
\end{eqnarray}
Thus the effective Hamiltonian is nothing but the $XXZ$ Heisenberg
chain, with anisotropy $J_{z}/J_{xy} \equiv \Delta =1/2$ in a
uniform and staggered longitudinal magnetic fields. It is worth
to notice that a similar problem has been studied intensively in
recent years \cite{CuBen_Affleck_Oshikawa,CuBen_Essler_Tsvelik,Mila_04,Kolezhuk_2004,Mah07}.

\subsection{The first critical field $H_{c_{1}}$ and the saturation field $H_{c_{2}}$}

The performed mapping allows to determine critical fields $H_{c_1}$
corresponding to the onset of magnetization in the system and the
saturation field $H_{c_2}$ \cite{Mila_98}. The easiest way to express $H_{c_1}$ and
$H_{c_2}$ in terms of ladder parameters is to perform the
Jordan-Wigner transformation which maps the problem onto a system of
interacting spinless fermions \cite{Luther_Peschel_75}:
\begin{eqnarray}
H_{sf} & = & t\sum_{n}(a_{n}^{+}a_{n+1} + h.c.) + V
\sum_{n}\rho_{n}\rho_{n+1}
\nonumber \\
&-& \sum_{n}\big[\mu_{0} + (-1)^{n}\mu_{1} \big]\rho_{n},
\label{Hamiltonian_SpFrm}
\end{eqnarray}
where $t=V=J_{\parallel}/2$, $\mu_{0} = h_{eff}^{0}+
J_{\parallel}/2$ and $\mu_{1}= h_{eff}^{1}$. The lowest critical
field $H_{c_1}$ corresponds to that value of the chemical
potential $\mu_{0c}$ for which the band of spinless fermions starts
to fill up. In this limit we can neglect the interaction term in Eq.
(\ref{Hamiltonian_SpFrm}) and obtain the model of free massive
particles with spectrum $E^{\pm}(k)=-\mu_{0} \pm (J_{\parallel}^{2} cos^{2}(k) + \mu_{1}^{2})^{1/2}$.
This gives $H_{c_1} = J_{\perp} -(J_{\parallel}^{2} + \delta^{2}J_{\perp}^{2})^{1/2}$.
A similar argument can be used to determine $H_{c_2}$. In the limit
of almost saturated magnetization, it is useful to make a
particle-hole transformation and estimate $H_{c_2}$ from the
condition where the transformed hole band starts to fill, what gives
$H_{c_2} = J_{\perp} +(J_{\parallel}^{2} + \delta^{2} J_{\perp}^{2})^{1/2}$.

\subsection{Magnetization plateau: $H_{c}^{\pm}$}

To determine parameters characterizing the magnetization plateaux
at $M=0.5M_{sat}$, we use the continuum-limit bosonization
treatment of the model (\ref{L_w_ARE_Effective_XXZ_Hamiltonian}).
To obtain the continuum version of the Hamiltonian
(\ref{L_w_ARE_Effective_XXZ_Hamiltonian}) we use the standard
bosonization expression of the spin operators \cite{GNT}
\bea \tau_{n}^{z}  &=&   \sqrt{\frac{K}{\pi}} \partial_x \phi \, +
 \, \frac{(-1)^n}{\pi} \sin(\sqrt{4\pi K}\phi) \, ,\label{bosforSz}\\
\tau_{n}^{+} &=&
\frac{e^{-i\sqrt{\pi/K}\theta}}{\sqrt{2\pi}}\left[\,
(-1)^{n}+\sin\sqrt{4\pi K}\phi \, \right],\label{bosforS+} \eea
where $\phi(x)$ and $\theta(x)$ are dual bosonic fields, $\partial_t
\phi = v_{s} \partial_x \theta $ and the spin-stiffness parameter
$K=\pi/2\left(\pi-\arccos\Delta \right)$.
Using (\ref{bosforSz})-(\ref{bosforS+}) we get the following
bosonized Hamiltonian
\begin{eqnarray}
H_{Bos}& =& \int dx \Big\{ \frac{v_{s}}{2}[(\partial_{x}\phi)^{2} +
(\partial_x\theta)^{2} ]
 - h_{eff}^{0} \sqrt{\frac{K}{\pi}}\partial_{x}\phi\nonumber\\
&+&  \frac{h_{eff}^{1}}{\pi a_{0}}
\sin(\sqrt{4\pi K}\phi)\Big\}\,
\label{L_w_ARE_Bosonized_Hamiltonian}
\end{eqnarray}
which is easily recognized as the standard Hamiltonian for the
commensurate-incommensurate phase transition
\cite{C_IC_transition} and the Bethe ansatz \cite{JNW_1984}. We
use these results to describe the magnetization plateau and
transitions from a gapped (plateau) to the gapless Luttinger
liquid phase.

Let us start our consideration from the case $h_{eff}^{0}=0$
corresponding to $H=J_{\perp} +J_{\parallel}/2$, where the
Hamiltonian (\ref{L_w_ARE_Bosonized_Hamiltonian}) reduces to the
quantum sine-Gordon (SG) model with a massive term $\sim
h_{eff}^{1}\sin(\sqrt{4\pi K}\phi)$. From the exact solution of
the SG model \cite{DHN} it is known that the excitation spectrum is
gapfull for  $0 < K < 2$. At $1 < K < 2$ the excitation spectrum
of the model consists of solitons and antisolitons with mass $M$,
while for $0 < K < 1$ the spectrum contains also
soliton-antisoliton bound states ("breathers"). However for $1/2 <
K < 1$ the soliton mass $M$ remains the lowest excitation energy
scale in the model. Below we will restrict our consideration by
limiting cases of the isotropic antiferromagnetic and
ferromagnetic ladder, corresponding to the case $\Delta=1/2$
($K=3/4$) and $\Delta=-1/2$ ($K=3/2$), respectively. Thus, in
both considered cases the effective bosonized Hamiltonian is in
the gapped regime with gap determined by the soliton mass $M$.
Moreover from the exact solution \cite{Zamolodchikov_95} we can
also get the exact scaling relation between the soliton mass
physical $M$ and it's bare value $m_{0}=\delta J_{\perp}$
\be M \simeq J_{\parallel} \left(\delta
J_{\perp}/J_{\parallel}\right)^{1/(2-K)} \, .
\label{SG-mass_Zamolodchikov} \ee
Thus at $h_{eff}^{0}=0$ the excitation spectrum of the system is
gapped. In the ground state the field $\phi$ is pinned in one of the
minima where $\la 0| \sin(\sqrt{4\pi K}\phi)|0 \ra =-1$, what corresponds (see Eq. (\ref{bosforSz}))
to a long-range antiferromagnetic  order in the ground state of the
effective Heisenberg chain, i.e. to a phase of the initial ladder
system, where  odd rungs have a dominant triplet character and even
rungs are predominantly singlets.

At $h_{eff}^{0} \neq 0$ the very presence of the gradient term in
the Hamiltonian (\ref{L_w_ARE_Bosonized_Hamiltonian}) makes it
necessary to consider the ground state of the sine-Gordon model in
sectors with nonzero topological charge. The effective chemical
potential $\sim h_{eff}^{0}\partial_{x}\phi$ tends to change the
number of particles in the ground state, what immediately implies
that the vacuum distribution of the field $\phi$ will be shifted
with respect of the minima which minimize the staggered part. This
competition between contributions of the smooth and staggered
components of the magnetic field is resolved as a continuous phase
transition from a gapped state at $|h_{eff}^{0}| < M$ to a gapless
Luttinger liquid phase at $|h_{eff}^{0}| > M$, where $M$ is the
soliton mass \cite{C_IC_transition}. This condition immediately
gives two additional critical values of the magnetic field
\begin{equation}
H_{c}^{\pm} = J_{\perp} + J_{\parallel}/2 \pm M, \label{Hc-pm}
\end{equation}
and respectively the width of the magnetization plateau given
by
\be H_{c}^{+}- H_{c}^{-} \simeq 2J_{\parallel}\left(\delta
J_{\perp}/J_{\parallel} \right)^{1/(2-K)} \, . \label{Platou} \ee
It is straightforward to get, that in the case of a ladder with
antiferromagnetic legs, the width of the magnetization plateau
scales as $\delta^{4/5}$ while in the case of a ladder with
ferromagnetic legs as $\delta^{2}$.

In order to investigate the detailed behavior of the ground state
magnetic phase diagram and to test the validity of the picture
obtained from continuum-limit bosonization treatment, below in this
paper we present results of numerical calculations using the Lanczos
method of exact diagonalizations for finite ladders.

\section{Numerical results} \label{sec3}

In this section, to explore the nature of the spectrum and the
quantum phase transition, we used Lanczos method to diagonalize
numerically ladders with length up to $L=14$.

\subsection{The Energy Gap}\label{subsec-3a}

First, we have computed the three lowest energy eigenvalues of
ladders with antiferromagnetic (ferromagnetic) legs
$J_{\parallel}=1.0 (-1.0)$ and different values of the rung
exchanges $J_{\perp}(n)$. To get the energies of the few lowest
eigenstates we consider ladders with periodic boundary
conditions on legs.

\begin{figure}
\centerline{\psfig{file=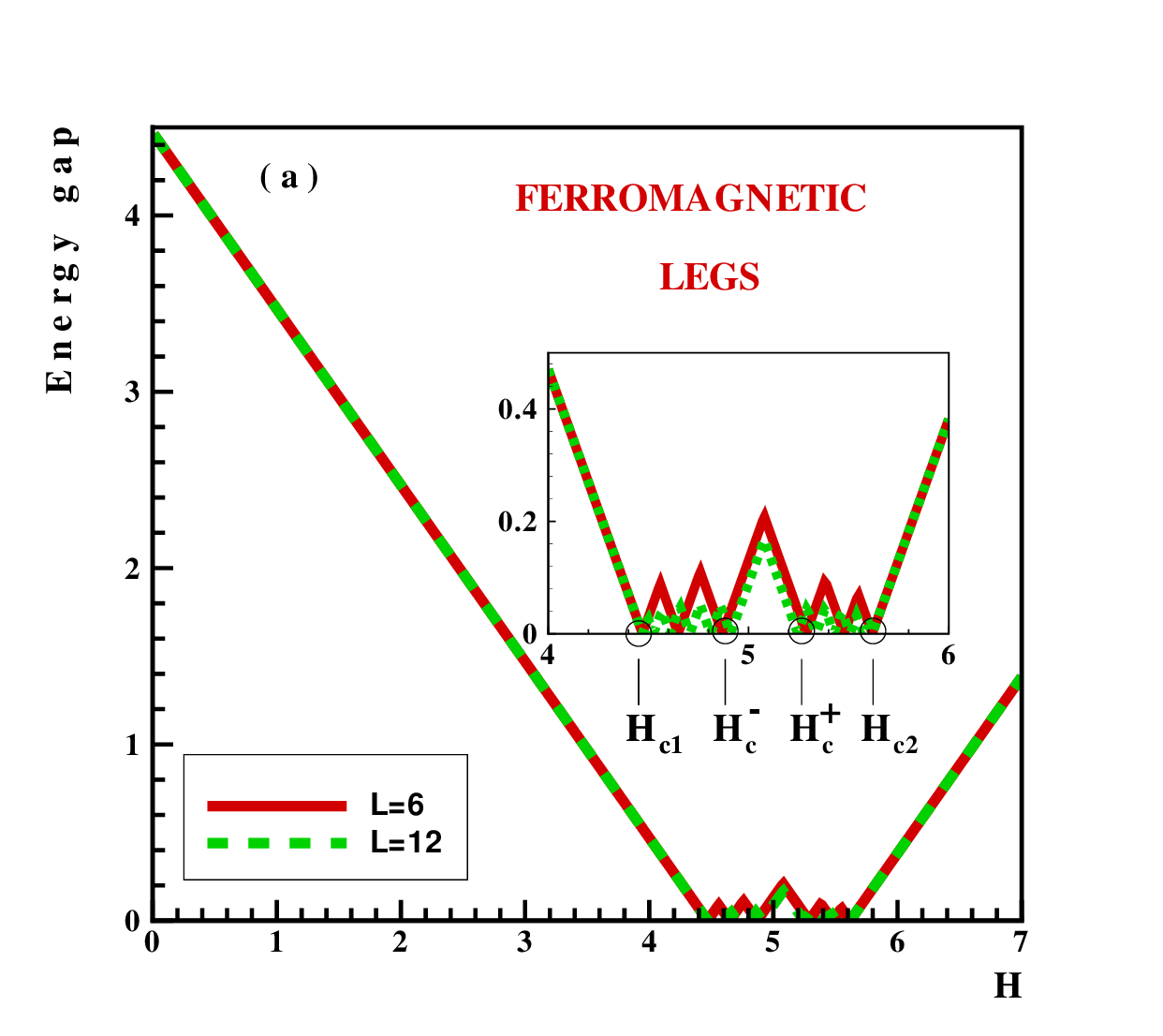,width=3.65in}}
\centerline{\psfig{file=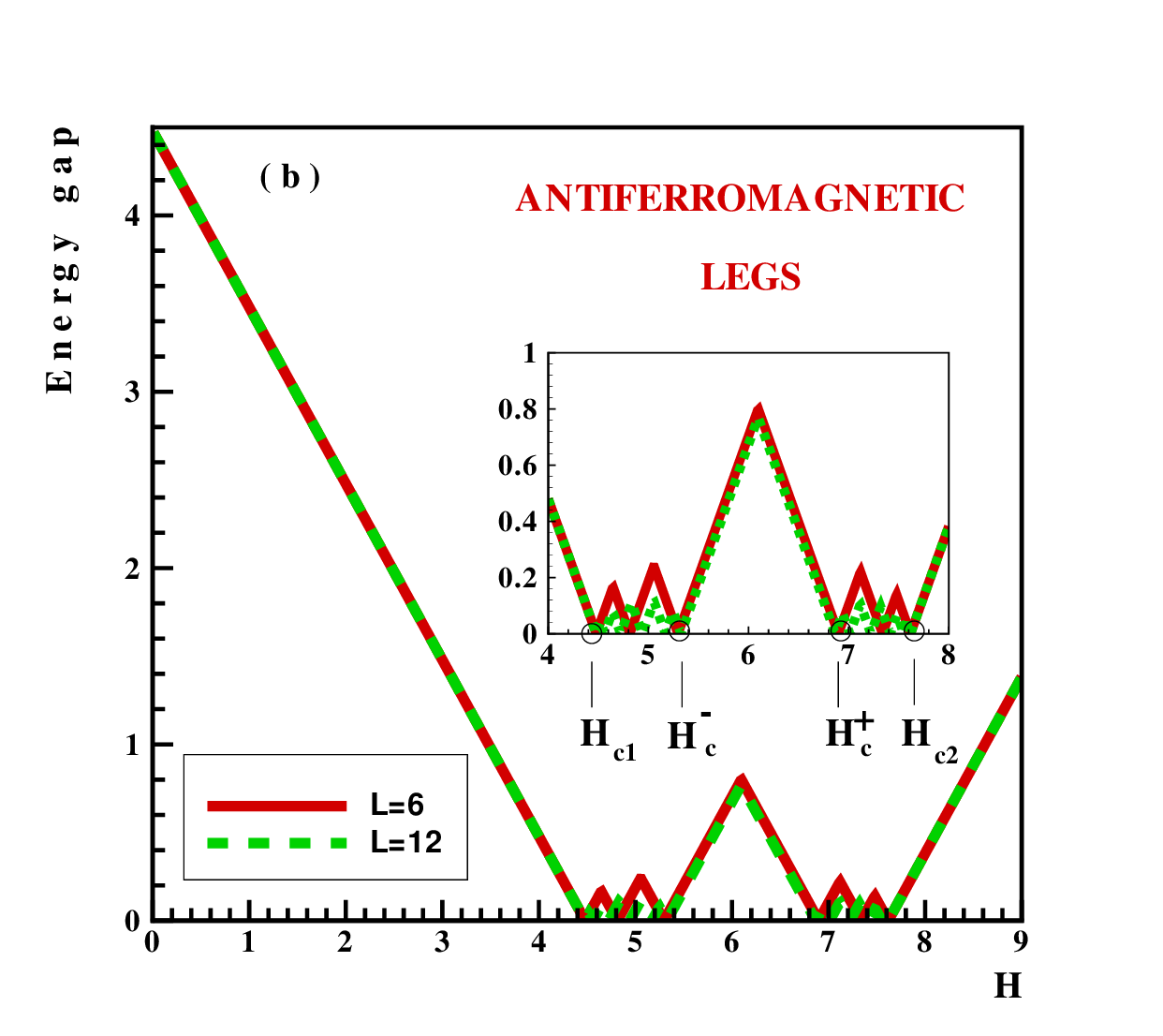,width=3.65in}}
\caption{Difference between the energy of the first excited state
and the ground state energy as a function of the magnetic field
$H$, for ladder with rung exchange $J_{\perp}=11/2$, $\delta=1/11$
and with (a) ferromagnetic legs $J_{\parallel}=-1.0$ (b)
antiferromagnetic legs $J_{\parallel}=1.0$, including different
ladder lengths $L=8, 10$. } \label{energy-gap}
\end{figure}

In Fig.\ref{energy-gap}, we present
results of these calculations for the rung exchanges
$J^{0}_{\perp}=11/2, \delta=1/11$ and ladder sizes $L=8,10$. We
define the excitation gap as a gap to the first excited state.
As it can be seen in Fig.\ref{energy-gap}, in the considered limit of strong
on-rung exchange, this difference is characterized by the indistinguishable (within the used numerical accuracy)
dependence on the ladder length and shows an universal linear decrease with increasing magnetic field.

At $H=0$ the spectrum model is gapped for all cases (ferromagnetic
and antiferromagnetic legs). For $H \neq 0$ the energy gap decreases
linearly with $H$ and vanishes at $H_{c_1}$. This is the first level
crossing between the ground-state energy and the first excited state
energy. The spectrum remains gapless for $H_{c_1}<H<H_{c}^{-}$ and
once again becomes gapped for $H>H_{c}^{-}$. The spin gap, which
appears at $H> H_{c}^{-}$, first increases with the external uniform
magnetic field, but then starts to decrease, and again vanishes at
$H_{c}^{+}$. With more increasing field $H>H_{c}^{+}$, the spectrum
remains gapless up to the critical saturation field $H_{c_2}$.
Finally, at $H>H_{c_2}$ the gap opens again and for a sufficiently
large field becomes proportional to $H$. In the regions
$H_{c_1}<H<H_{c}^{-}$ and $H_{c^{+}}<H<H_{c_{2}}$, the two lowest
states cross each other $N/2$ times, which causes gapless regimes in
the thermodynamic limit $N\longrightarrow \infty$. These level
crossings lead to incommensurate effects that manifest themselves in
the oscillatory behavior of the response functions.

\begin{figure}[tbp]
\centerline{\psfig{file=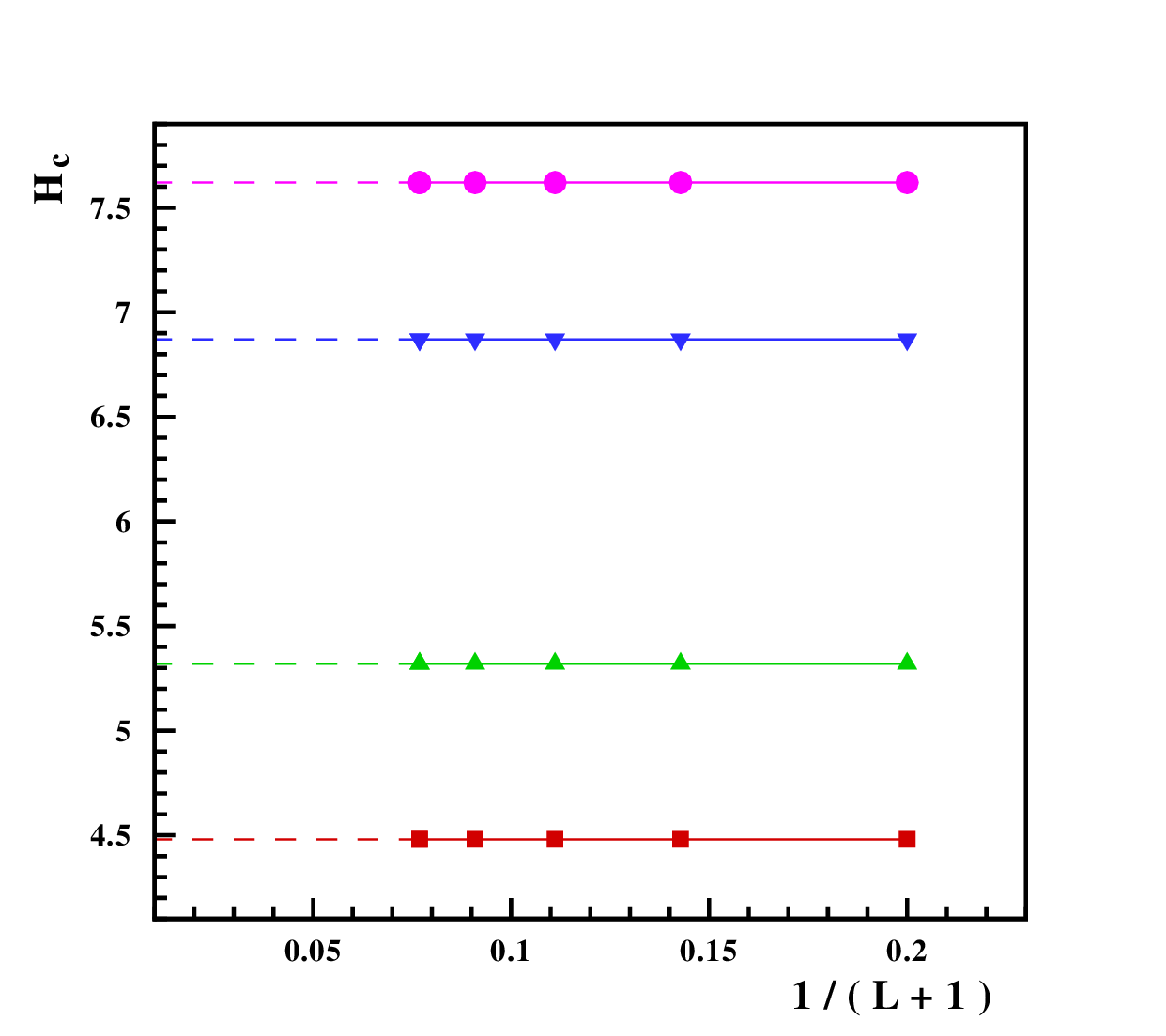,width=3.65in}}
\caption{ Fixed points $H_{c}(L, L +2)$ of the PRG equation for
the energy gap, plotted versus $1/(L+1)$, for $J_{\parallel}=1.0$,
$J_{\perp}=11/2, \delta=1/11$ and different ladder lengths $L=4,
6, 8, 10, 12, 14$. The values of $H_{c}(L, L+2)$ for five pairs of
system sizes $(L, L+2)= (4, 6), (6, 8), (8, 10), (10, 12)$, and
$(12, 14)$ are represented in the figure.} \label{PRG}
\end{figure}

Let us now comment the numerical method which we use to extract
critical values of the magnetic field in the thermodynamic limit ($L
\rightarrow \infty$) from the finite ladder exact diagonalizations
data. In an infinite system the critical point corresponds to a
value of the magnetic field at which the energy gap vanishes. Since,
in the case of finite systems the gap remains finite, we use the
phenomenological renormalization group (PRG) method (see for details
Ref.44 ) to obtain values of the critical fields. The
phenomenological renormalization group equation for the energy gap
$G(L, H)$ is written as
\begin{eqnarray}
(L+2) G(L+2,H')=L G(L, H)\, . \label{prg}
\end{eqnarray}
At the quantum critical point, $L G(L, H)$ should be size
independent for large enough systems in which the contribution
from irrelevant operators is negligible. This allows to determine
critical points very accurately. We define $H_{c}(L, L+2)$ as
$L$-dependent fixed point of PRG equation and then it is
extrapolated to the thermodynamic limit in order to estimate
$H_{c}$. At the first step we plot the curve $LG(L,H)$ versus $H$
for system sizes $L$ and $L+2$. These curves cross at a certain
value $H_{c}(L, L+2)$ which is determined as a finite size
critical point. The thermodynamic critical point $H_{c}$ is
obtained by extrapolating $H_{c}(L, L+2)$ to $L\longrightarrow
\infty$. Figure~\ref{PRG} shows the extrapolation procedure of the
transition points for the ladder with antiferromagnetic legs and
rung-exchange parameters $J_{\perp}=11/2$ and $\delta=1/11$.
Obtained by this method values of critical are
\begin{eqnarray}
H_{c_1} &=& 4.48 \pm 0.01,~~~~~H_{c_2} = 7.62 \pm 0.01,\\
H_{c}^{-}&=& 5.32 \pm 0.01,~~~~~H_{c}^{+}=6.87\pm0.01\, .
\end{eqnarray}
Analogously in the case of the ladder with ferromagnetic legs we obtain
the following set of critical parameters
\begin{eqnarray}
H_{c_1} & = & 4.47\pm0.01,~~~~~H_{c_2} = 5.62 \pm 0.01,\\
H_{c}^{-} &= & 4.89\pm0.01,~~~~~H_{c}^{+}= 5.26\pm0.01\, .
\end{eqnarray}
One can easily check that the obtained values for critical field
obtained from the finite ladder studies are very close to the values
predicted analytically.

\subsection{Magnetization curve}\label{subsec-3b}

\begin{figure}
\centerline{\psfig{file=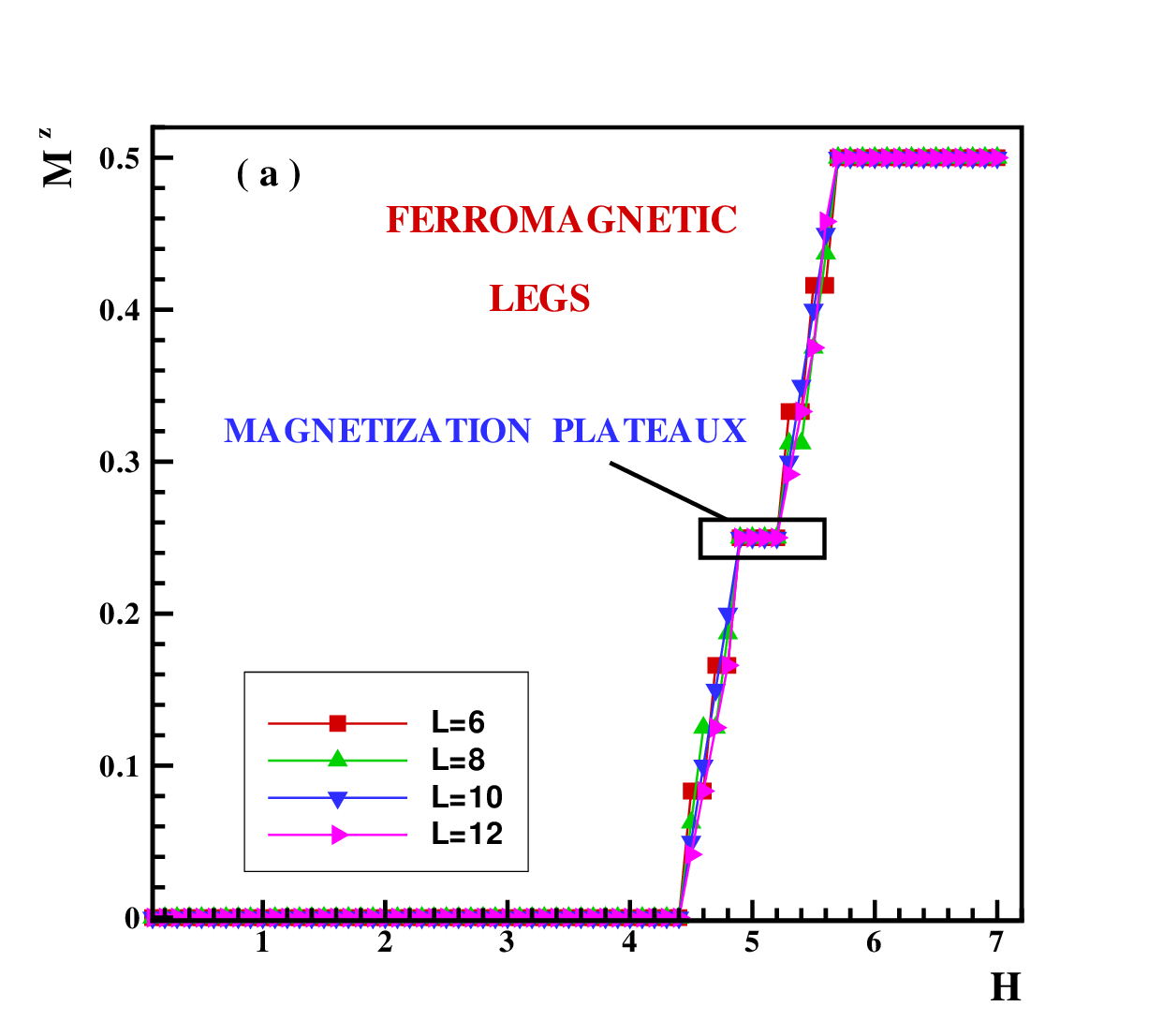,width=3.65in}}
\centerline{\psfig{file=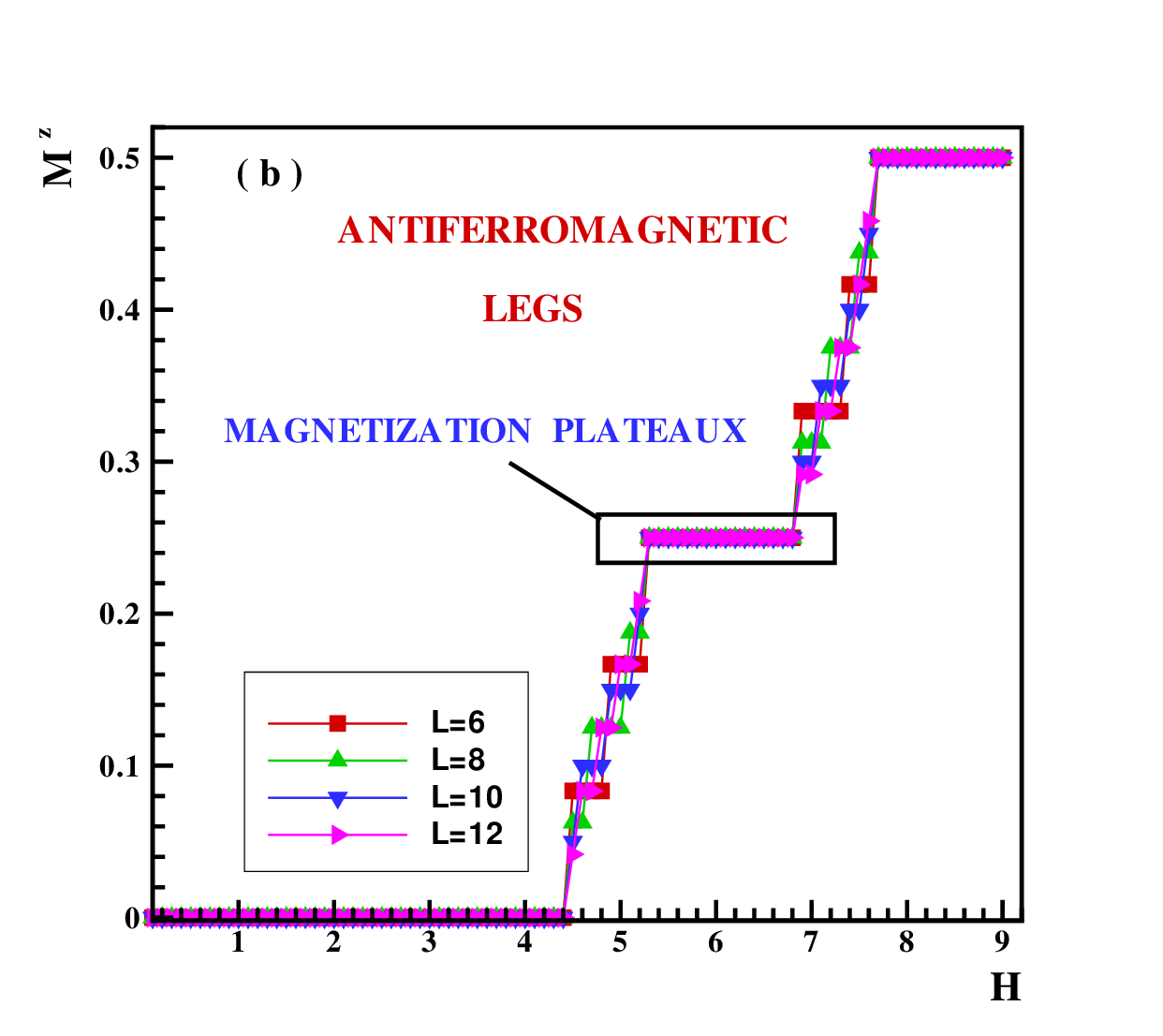,width=3.65in}}
\caption{a. The magnetization along field $M^z$ as a function of
the applied magnetic field $H$ for (a) ferromagnetic ladder
$J_{\parallel}=-1.0$ (b) antiferromagnetic ladder
$J_{\parallel}=1.0$, including different ladder lengths $L=6, 8,
10, 12$.  } \label{Magnetization}
\end{figure}

To study the magnetic order of the ground state of the system, we
start with the magnetization process. First, we have implemented
the Lanczos algorithm on finite ladders to calculate the lowest
eigenstate. The magnetization along the field axis is defined as
\begin{eqnarray}
M^{z}=\frac{1}{L}\sum_{n=1}^{L}\langle
Gs|\left(S_{1,n}^{z}+S_{2,n}^{z}\right)|Gs\rangle,
\label{magnetization}
\end{eqnarray}
where the notation $\langle Gs|...|Gs\rangle$ represent the ground
state expectation value. In Fig.~\ref{Magnetization}, we have
plotted $M^{z}$ as a function of the magnetic field $H$, and for
a ladder with (a) ferromagnetic  and (b)antiferromagnetic  legs
and with rung exchange parameters $J_{\perp}=11/2, \delta=1/11$
and different lengths $L=6, 8, 10, 12$ .

As it is clearly seen in Fig.~\ref{Magnetization} besides the
standard rung-singlet and saturation plateaus at $H<H_{c_1}$ and
$H>H_{c_2}$ respectively observed in the case of ladders with uniform rung exchange
 \cite{Poilblanc_96,Chitra_Giamarchi_97}, we observe a plateau at
$M=0.5M_{sat}$. Observed oscillations of the magnetization at
$H_{c_1}< H <H_{c}^{-}$ and $H_{c}^{+} < H< H_{c_2}$ result from the
level crossing between the ground and the first excited states of
this model in the gapless phases.  Usually, in order to
give an estimation of the width of magnetization plateau in the
thermodynamic limit, the size scaling of its width is performed
\cite{Honecker00}. We also perform the similar analysis for selected
values of the exchange parameters and found that the size of the
plateau interpolates to finite value with $L \rightarrow \infty$.

\begin{figure}
\centerline{\psfig{file=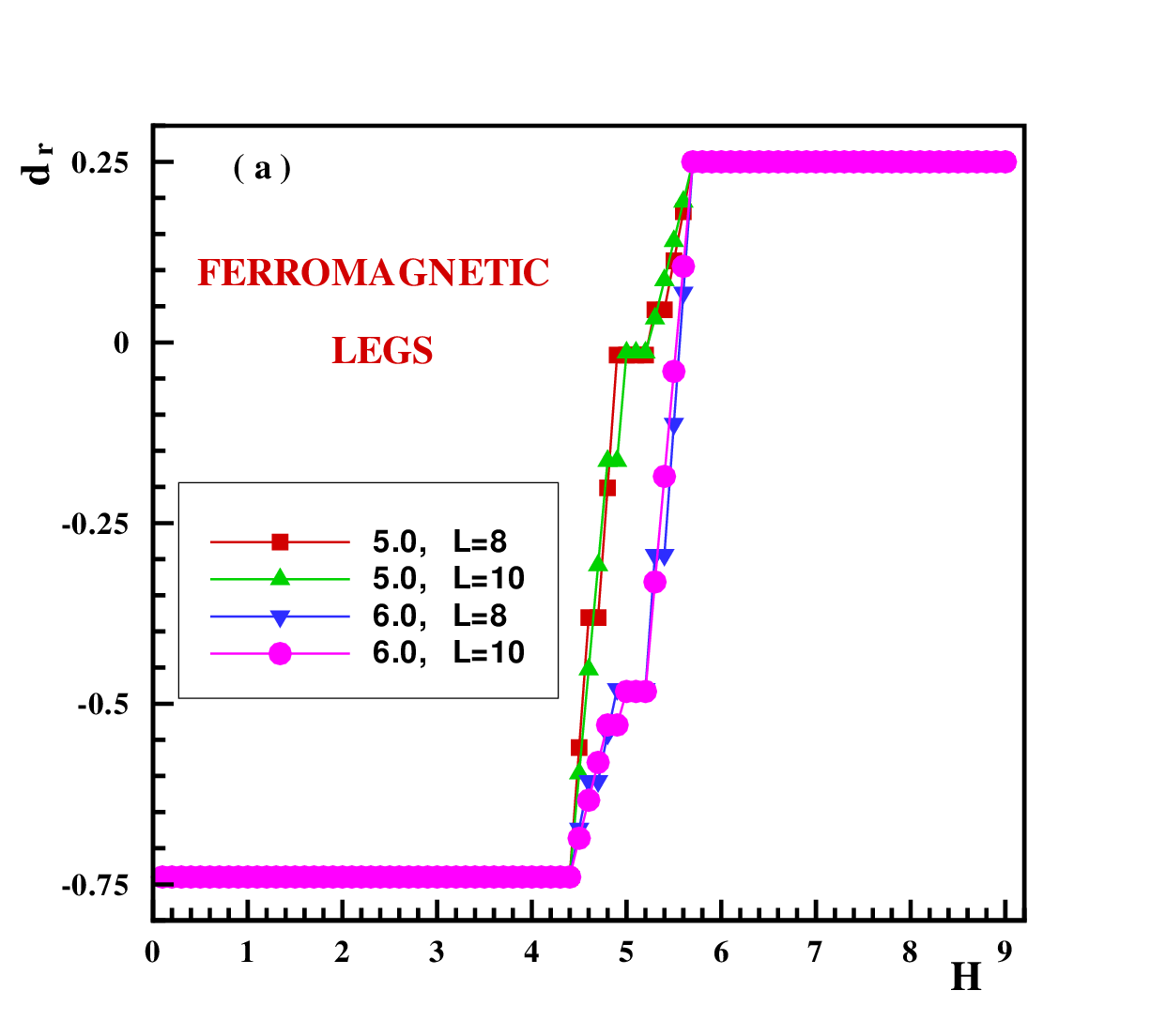,width=3.65in}}
\centerline{\psfig{file=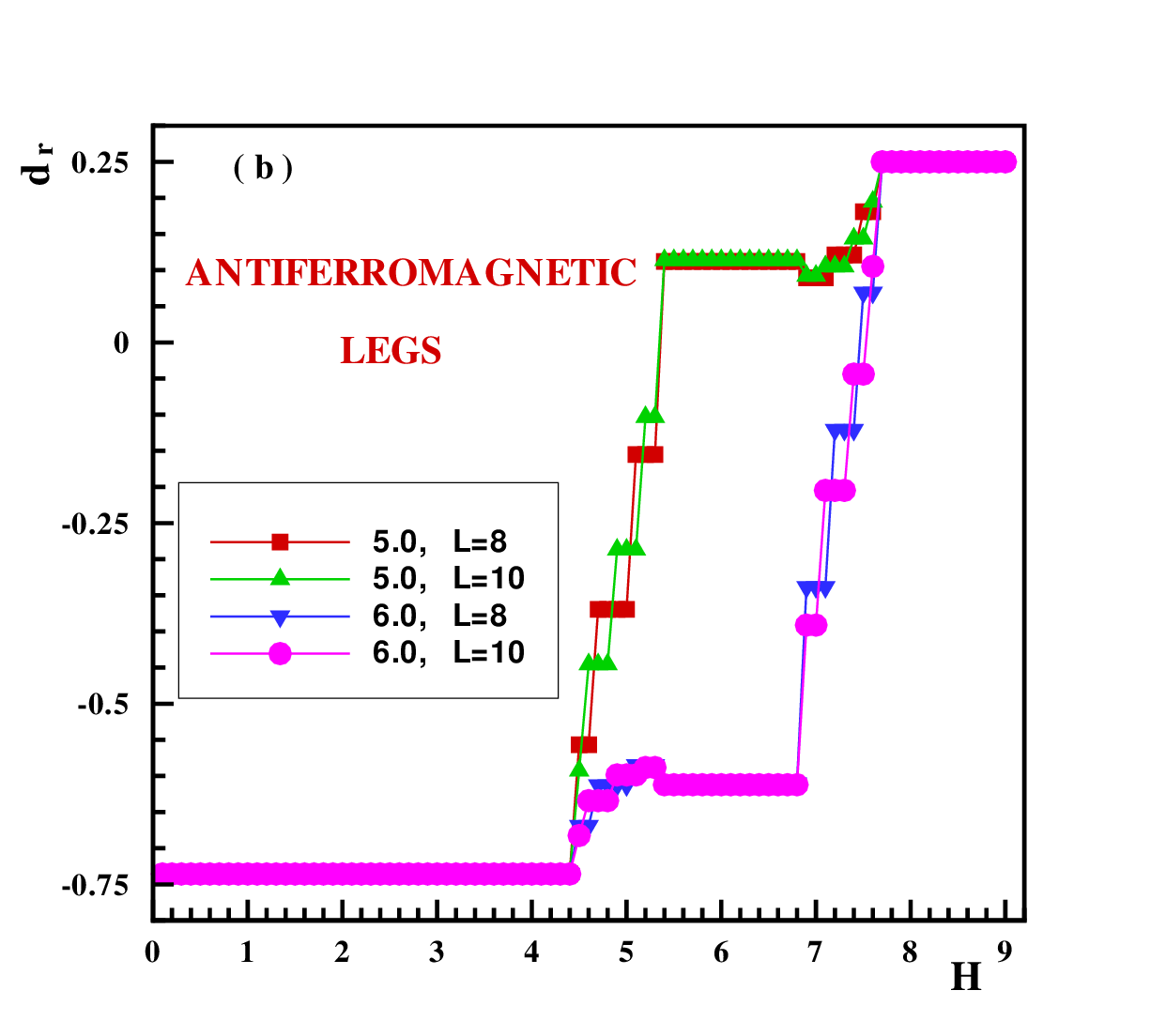,width=3.65in}}
\caption{The on-rung spin correlation functions for odd ($J_{\bot}^{-}=5.0$)
and even ($J_{\bot}^{+}=6.0$) rungs  as a function of the applied field $H$
for $L=8, 10$ lengths with rung exchange parameters $J_{\perp}=11/2,
\delta=1/11$ with (a) ferromagnetic ($J_{\parallel}=-1.0)$  and
(b) antiferromagnetic legs ($J_{\parallel}=1.0$) .}
\label{dimerization}
\end{figure}

An additional insight into the nature of different phases can be
obtained studying the intra-rung correlations. We define the
on-rung spin correlation function (rung dimerization order
parameter) for even and add sites, as
\begin{eqnarray}
d^{e}_r=\frac{2}{L}\sum_{m}\langle Gs| {\bS}_{1, 2m}\cdot{\bS}_{2,
2m}|Gs\rangle  \label{dimerization_E}
\end{eqnarray}
and
\begin{eqnarray}
d^{o}_r=\frac{2}{L}\sum_{n}\langle Gs| {\bS}_{1, 2m+1}\cdot {\bS}_{2,
2m+1}|Gs\rangle \, , \label{dimerization_O}
\end{eqnarray}
taking sum over even or odd sites, respectively. In
Fig.~\ref{dimerization} we have plotted the $d^{e}_r$ and
$d^{o}_r$ as a function of the magnetic field $H$ for the ladder of
lengths $L=8, 10$ with (a) ferromagnetic  and (b) antiferromagnetic
legs and the rung exchange parameters $J_{\perp}=11/2,
\delta=1/11$. As it is seen from this figure, at $H<H_{c_1}$ spins
on all rungs are in a singlet state $d_{r}^{e}=d_{r}^{o} \simeq
-0.75$, while at $H>H_{c_2}$, $d_{r}$ is slightly less than the
saturation value $d_{r}^{e}=d_{r}^{o}\sim 1/4$ and the
ferromagnetic long-range order along the field axis is present.
Deviation from the saturation values $-3/4$ and $1/4$ reflects
the effect of quantum fluctuations. However, in the considered
case of strong rung-exchanges and high critical fields quantum
fluctuations are substantially suppressed and calculated averages
of on-rung spin correlations are very close to their nominal
values.

On the other hand, for intermediate values of the magnetic field,
at $H_{c_1}<H<H_{c_2}$ the data presented in Fig.~\ref{dimerization}
gives us a possibility to trace the mechanism of
singlet-pair melting with increasing magnetic field. As it follows from Fig.~\ref{dimerization}
at $H$ slightly above $H_{c_1}$ spin singlets pairs start to melt in all rungs
simultaneously and almost with the same intensity. This effect is
more profound in the case of ladder with ferromagnetic legs (Fig.
~\ref{dimerization}.a) where the on-rung spin correlation functions
increase almost parallel for even and odd rungs. With further
increase of $H$ melting of weak rungs gets more intensive,
however at $H=H^{-}_{c}$ the process of melting stops. As it is
seen in Fig.~\ref{dimerization}.b weak rungs are polarized,
however their polarization is far from the saturation value
$d^{o}_r\simeq 0.1$, while the strong rungs still manifest strong
on-site singlet features with $d^{e}_r \simeq -0.62$. This effect
is much weaker in the case of ferromagnetic legs in accordance
with previous results.  Moratorium on melting stops at
$H=H^{+}_{c}$, however for $H>H^{+}_{c}$ strong rungs start to
melt more intensively while the polarization of weak rungs
increases slowly. Finally at $H=H_{c_2}$ both subsystems of rungs
achieve an identical, almost fully polarized state. Note, that
the almost symmetric fluctuations in on-rung correlations,
increase in $d^{e}_r$ at $H \leq H^{-}_{c}$
decrease in $d^{o}_r$ at
$H \geq H^{+}_{c}$ reflect the enhanced role of quantum
fluctuations in the vicinity of quantum critical points.

\begin{figure}[tbp]
\centerline{\psfig{file=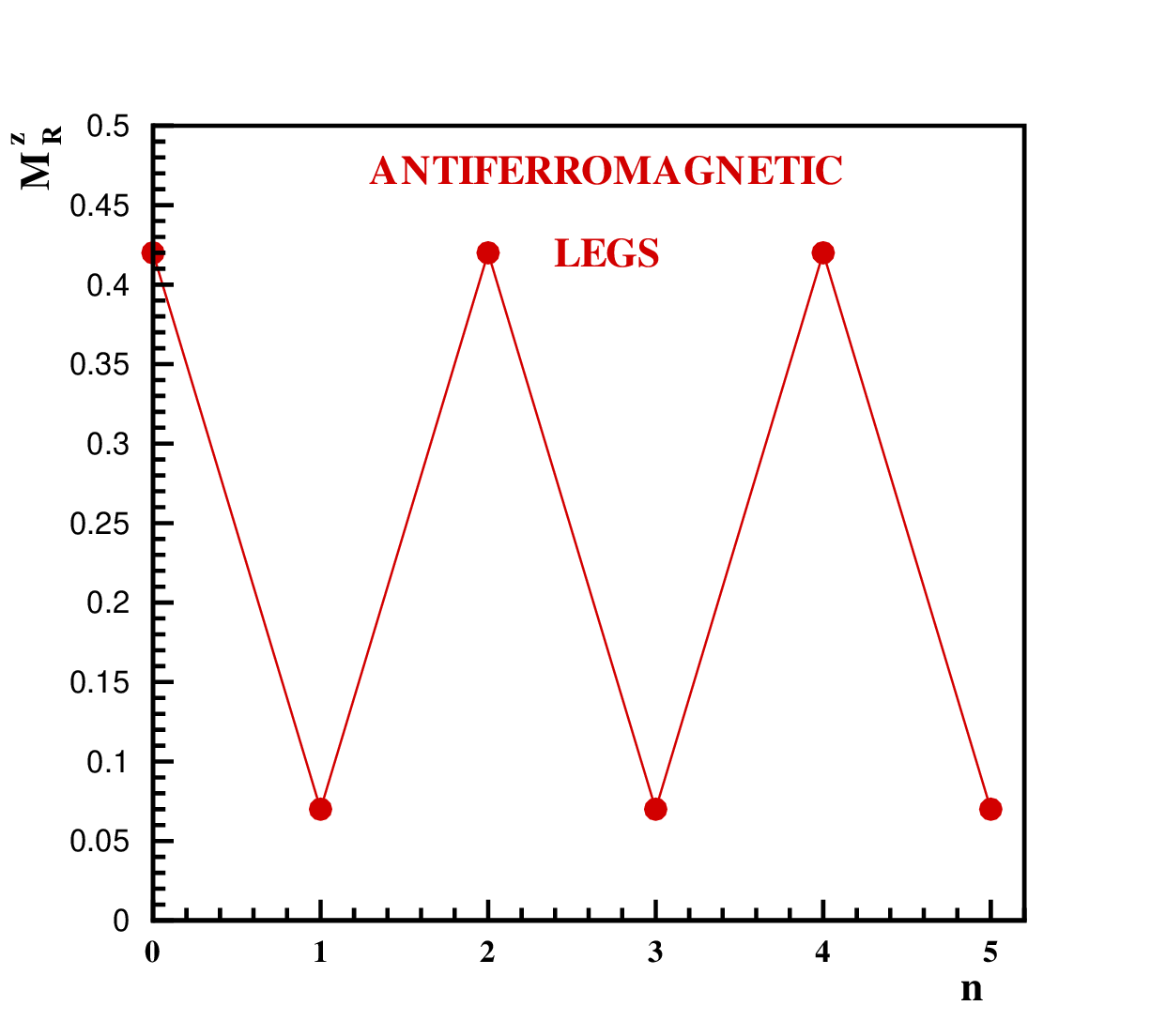,width=3.65in}}
\caption{The spin distribution in the GS as a function of the rung
number "n" for magnetization corresponding to plateau at
$M^{z}=0.5M_{sat}^{z}$ and for rung exchanges $J_{\perp}^{-}=5.0$
and $J_{\perp}^{+}=6.0$. } \label{Spin.Distribution}
\end{figure}

To complete our description of the phase at magnetization plateau
with $M=0.5M_{sat}$  we have calculated the rung-spin distribution
in the ground state
\begin{eqnarray}
M^{z}_{n}=\frac{1}{2}\langle
Gs|\left(S_{n,1}^{z}+S_{n,2}^{z}\right)|Gs\rangle.
\label{distribution}
\end{eqnarray}
In Fig. \ref{Spin.Distribution} we have plotted the spin
distribution in the ground state of a ladder with rung-exchange
parameters $J_{\perp}=11/2,\delta=1/11$ and antiferromagnetic legs
as a function of the rung number "n" for a value of the magnetic
field corresponding to the plateau at $M^{z}=0.5M_{sat}^{z}$. To
obtain an accurate estimate of the function $M^{z}_{n}$, we have
calculated them from the Eq.(~\ref{distribution}) for system
sizes of $L=6, 8, 10, 12, 14$. The thermodynamic limit ($L
\rightarrow \infty$) of the finite size results are obtained by
extrapolation method and used for plotting. As we observe the
system shows a well pronounced modulation of the on-rung
magnetization, where magnetization on odd rung  is larger than on
even rungs. This distribution remains almost unchanged
within the plateau for $H_{c}^{-}<H<H_{c}^{+}$.

\subsection{Scaling properties of the magnetization  plateau}

\begin{figure}
\centerline{\psfig{file=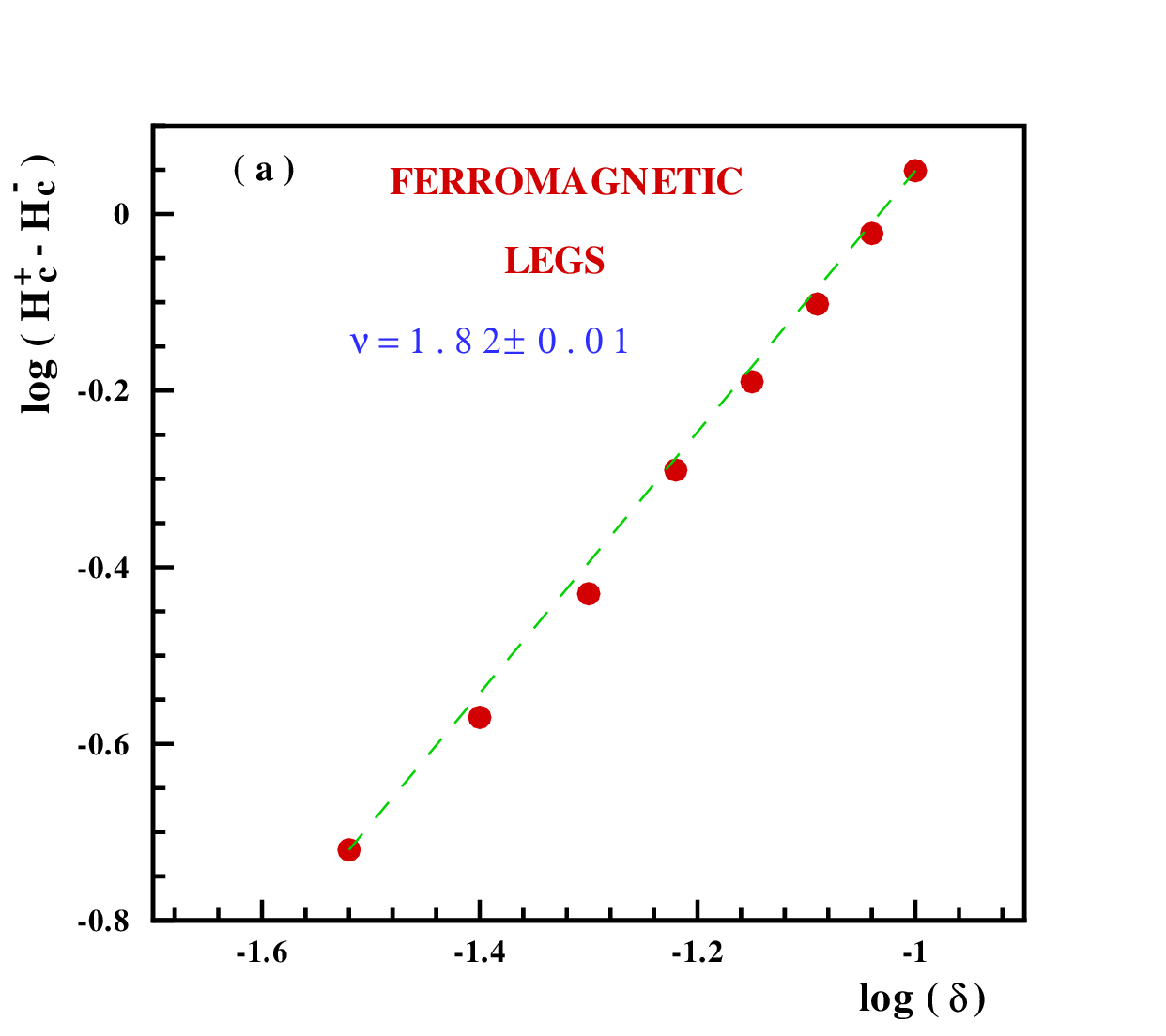,width=3.65in}}
\centerline{\psfig{file=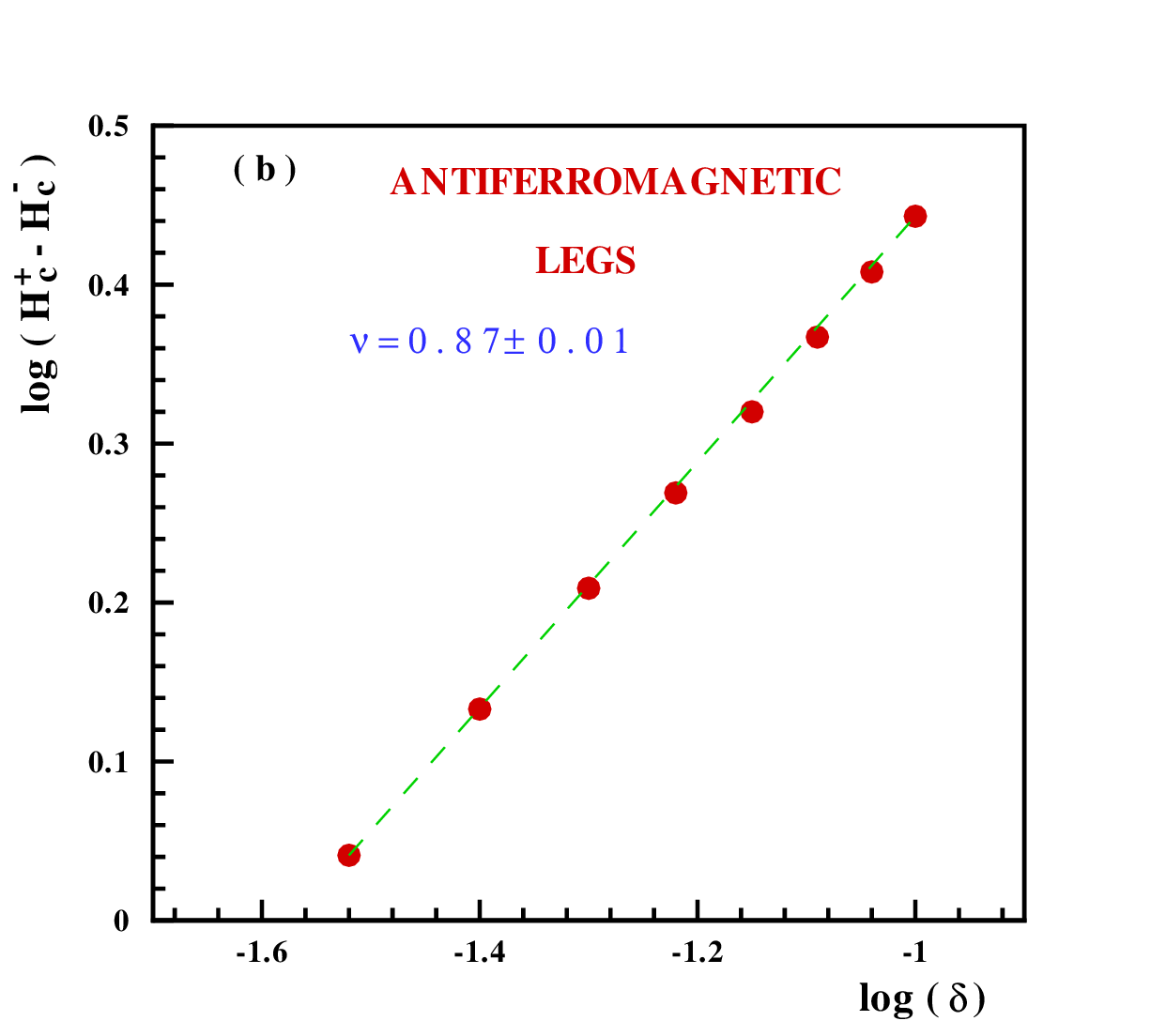,width=3.65in}}
\caption{Width of the magnetization plateau as a function of
parameter $\delta$ for $0.01<\delta<0.1$ in the case of ladder
with $J_{\bot}=10.0$ and, (a) ferromagnetic
($J_{\parallel}=-1.0$)  and (b) antiferromagnetic
($J_{\parallel}=1.0$) legs.} \label{plateau.index}
\end{figure}

To find an accurate estimate on the critical exponent
characterizing width of the magnetization plateau on the
parameter $\delta$  we have computed the critical fields
$H_{c}^{\pm}$ for the finite ladder systems with $L =6, 8, 10, 12,
14$, $J_{\perp}=10$ and different values of the parameter
$\delta$ ($0.01<\delta<0.1$) and obtain their extrapolated values
corresponding to the thermodynamic limit $L \rightarrow \infty$.
In Fig.~\ref{plateau.index} we have plotted the log-log plot of the plateau width
versus $\delta$. Calculations has been performed both in the case
of ladder with antiferromagnetic and ferromagnetic legs.  We
found that the best fit to our data (using the equation
$H_{c}^{+}-H_{c}^{-} \sim \delta^{\nu}$) yields $\nu=1.82\pm
0.02$ and $\nu=0.87\pm 0.01$ for the ladders with ferromagnetic
and antiferromagnetic legs respectively.

\section{Conclusion}

In this paper we have studied the elementary excitations and the
magnetic ground state phase diagram of a spin $S=1/2$ two-leg
ladder with alternating rung-exchange $J_{\perp}(n) =
J_{\perp}\left[1 + (-1)^{n} \delta \right]$ using the continuum
limit bosoni\-zation studies and the Lanczos method of numerical
diagonalizations for ladders up to $L=14$.  We have shown that the
rung-exchange alternation leads to generation of a gap
in the excitation spectrum of the system at magnetization equal
to the half of its saturation value. As a result of this new energy scale formation
the magnetization curve of the system $M(H)$ exhibits a plateau at $M=0.5M_{sat}$. The width of
the plateau, is proportional to the excitation gap and scales as
$\delta^{\nu}$, where critical exponent $\nu =0.87\pm0.01$ in the
case of a ladder with isotropic antiferromagnetic legs and $\nu
=1.82\pm0.01 $ in the case of a ladder with isotropic
ferromagnetic legs. We have also calculated the magnetic field dependence of the on-rung
spin-spin correlation functions. Comparison of these data for weak
and strong rungs gave us an excellent description of the dynamics of
the magnetization process in the case of ladder with non-equal
rungs.

In a standard way we estimate the magnetic condensation energy as
$E_{mag}(\delta)-E_{mag}(0) \sim -\delta^{2\nu}$. In the harmonic approximation
the lattice deformation energy (per rung) can be estimated as $E_{def} \sim
\delta^{2}$. Therefore we can conclude, that in the case of antiferromagnetic ladder,
where  $2\nu = 1.64 < 2$, the spontaneous appearance of an alternating rung exchange
as a spin-Peierls instability during the magnetization process at $M=0.5M_{sat}$ is
possible.

\section{Acknowledgments}

It is our pleasure to thank  T. Vekua for fruitful discussions. This
work has been supported by the GNSF through the grant No.
ST06/4-018. GIJ also acknowledges support through the research program of
the SFB 608 funded by the DFG.

\vspace{0.3cm}

\end{document}